\documentclass[a4paper,11pt]{article}
\pdfoutput=1 

\usepackage{jheppub} 

\usepackage[T1]{fontenc} 
\usepackage{float}
\usepackage{epstopdf}
\usepackage{amsmath}
\makeatletter
\gdef\@fpheader{}
\makeatother

\title{\boldmath Lovelock black holes with double-logarithmic electrodynamics source}

\author{Askar Ali} 
\affiliation[1]{Department of Mathematics, Quaid-i-Azam University, Islamabad, Pakistan}

\affiliation[2]{Department of Sciences and Humanities, National University of Computer and Emerging Sciences, Peshawar 25000, Pakistan}

\emailAdd{askarali@math.qau.edu.pk}

\abstract{This work examines the magnetized black holes of Lovelock gravity in the presence of double-logarithmic electrodynamics. In this context, the Lovelock polynomial is found and the accompanying thermodynamic quantities, such as mass, entropy, Hawking temperature, and heat capacity are determined. This new model of nonlinear electrodynamics is used to calculate the black hole solutions of Einstein, Gauss-Bonnet and third order Lovelock gravities as well. The impacts of the double-logarithmic electromagnetic field on the black hole thermodynamics in these particular theories are examined and the regions of horizon radius that correspond to the local thermodynamic stability are highlighted.
\vspace{80 mm}
}

\notoc 

\begin{document}
\maketitle
\flushbottom 


\section{Introduction}
\label{sec:intro}
The most successful theory which illustrates our Universe at low energy scales is the well-known Einstein's theory of gravity \cite{1}. However, one cannot consider it a valid theory at higher energies comparable to the Planck scale. Therefore, it is necessary to incorporate the modifications of this theory. It is eminent that string theory and brane cosmology strongly predict the existence of higher dimensions \cite{2,2a,3,3a,3b}. Hence, the higher dimensional probing of gravitational fields seems to be very important. The most natural extension of Einstein's gravity in higher dimensions is the Lovelock gravity \cite{4}. This theory modifies Einstein's gravity with higher curvature terms and yields the second-order equations of motion in extra dimensions. Note that, the essential idea is to take the gravitational action as a sum of dimensionally continued Euler densities of all dimensions below $d$. Due to this, the $k$-th order term is either topological or equal to zero below spacetime dimension $d=2k+1$. The quantization of this theory at the linear level is free of ghosts \cite{5}. Thus, the Lovelock gravity can be considered a natural model for describing the effects of higher curvature terms and extra dimensions on gravitational physics. Another interesting feature of Lovelock gravity arises from its relationship with low energy effective string theory \cite{5,5a}. Many static spherically symmetric black holes of Lovelock gravity were studied in Refs. \cite{5b,5c,5d,5e,5f,5g,5h,5i,5j}. 

 The shortcomings of Maxwell's theory motivate one to study nonlinear electrodynamics (NLED) theories. Among these problems, one is the appearance of infinite self-energy of a charged particle. To tackle this difficulty Born and Infeld formulated an interesting model of NLED \cite{81}. The action associated with the Born-Infeld (BI) electrodynamics can also arise in the low energy limit of heterotic string theory \cite{82,83}. The impacts of the nonlinear electromagnetic fields and higher curvature gravities could not be avoided because it is anticipated that the gravitational and electromagnetic fields close to charged gravitating objects will be quite strong. Hence, Maxwell's electrodynamics and Einstein's gravity are, respectively, the approximations of NLED and higher curvature gravities in the weak field limit. Among the modifications of Einstein-Maxwell theory, the model of Lovelock gravity with NLED sources could be more interesting because it gives a ghost-free theory in the presence of a divergence-free nonlinear electromagnetic field. In addition to BI electrodynamics, other useful models have also been formulated \cite{83a,84,85,86,87,88,88a}. These models, like the BI formalism, are also particularly beneficial in studies of the black hole's properties, dark energy, and inflation. Moreover, these models include a few different characteristics as well. For example, in contrast to BI electrodynamics, the metric function for the interpretation of electrically charged gravitating objects in the presence of exponential \cite{86} and double-logarithmic electromagnetic \cite{88} sources can only be obtained in terms of integral functions. In a similar vein, the causality principle cannot be guaranteed by the BI formalism, although in the exponential, double-logarithmic and arcsine \cite{88a} models, both the causality and unitarity principles are not violated. Furthermore, the Heisenberg-Euler effective Lagrangian \cite{83a} which induces the quantum corrections in quantum electrodynamics causes the vacuum birefringence effect. Although vacuum birefringence is not a reality in the BI model, it does have an impact on exponential and double-logarithmic models.        
 
  As it leads to the regular black hole solution, NLED has also recently attained considerable interest. The nonlinear electrically charged regular black hole solution of Einstein's gravity has been derived in this study \cite{89}. For example, the metric function of the Bardeen black hole in the context of NLED is found in Ref. \cite{90}. Many exact black hole solutions in various gravity theories sourced by NLED models were derived in the literature \cite{98,99,100,101,102,103,104,105,106,107,108,109,110,111,112,113,114,115,116,117,118,119,120,121,121A,121B,122,123,123a1,123a2,123a3,123a4}. Recently, the impacts of double-logarithmic electromagnetic field on the dimensionally continued scalar hairy black holes \cite{124} and rotating Lovelock black branes \cite{124a} have also been examined. The objective of this paper is to study the effects of NLED on the properties of Lovelock black holes and their thermodynamics. Therefore, I used the model of double-logarithmic electrodynamics \cite{88} as a matter source and investigate the charged black holes of Lovelock gravity. This model of NLED is dependent on both Lorentz invariants, i.e., $\mathcal{P}=F_{\mu\nu}F^{\mu\nu}$ and $\mathcal{G}=F_{\mu\nu}\tilde{F}^{\mu\nu}$ and on the nonlinearity parameter $\beta$ as well. Due to the difficulty in the determination of dyonic and electric black hole solutions in closed form, we have only studied magnetized Lovelock black holes by taking the second invariant $\mathcal{G}$ equal to zero. 
  
   Analyzing the AdS/CFT correspondence is a key factor in the study of higher dimensional black holes with a negative cosmological constant. The thermodynamic properties of AdS black holes could provide profound insights about the phase structures of strong't Hooft coupling CFTs if this principle is taken into consideration. Under a specific temperature, it is demonstrated that there does not exist any AdS black hole solution, and that the well-known Hawking-Page phase transition can take place between stable large black holes and thermal gas in AdS space \cite{138}. Additionally, it is concluded that in a canonical ensemble, charged black hole systems cause the smaller/larger black hole (SB/LB) phase transition \cite{138a}. This demonstrates how the black hole system and Van der Waals fluid are quite similar in behavior. Along with the SB/LB phase transition, new phenomena that are analogous to the ``everyday thermodynamics'' of simple substances are also covered. Examples include the reentrant phase transitions of multicomponent liquids, multiple solid/liquid/gas (S/L/G) phase transitions, and Van der Waal type L/G  phase transitions. It was determined that the six or higher dimensional single spinning Kerr black hole with negative cosmological constant exhibits the peculiar behavior of LB/SB/LB phase transitions \cite{138b}. These reentrant phase transitions were also appeared in the four-dimensional AdS BI black holes \cite{138c}. Nevertheless, the higher dimensional version of these black hole do not exhibit the occurrence of these transitions. Recently, it has been noticed that the doubly-spinning Myers-Perry black holes of Einstein's theory may also experience similar transitions \cite{138d}. Similar to this, the reentrant phase transitions, triple points, and phase diagrams associated with charged black holes of Gauss-Bonnet gravity were explored in Ref. \cite{138e,138e1}. Given the importance of the thermodynamic properties of AdS black holes, we have also looked into how a double-logarithmic electromagnetic field affects the thermodynamic stability of different Lovelock black holes.      
 
This paper is organized as follows. In section 2, I investigate the new black hole solutions of Lovelock gravity derived from double-logarithmic electrodynamics. The Lovelock polynomial which gives the black hole solutions is developed and the important thermodynamic quantities related to these black holes are calculated. Section 3 covers the investigation of Einsteinian black holes and their thermodynamics. Similarly, sections 4 and 5 contain the investigation of magnetized Gauss-Bonnet and third order Lovelock black holes, respectively. Finally, I complete a paper with some closing remarks.

\section{Lovelock black holes and double-logarithmic electrodynamics} 

The $d$-dimensional action describing Lovelock gravity supported by an electromagnetic field can be defined as
\begin{equation}
\mathcal{I}_g=\mathcal{I}_L+\mathcal{I}_m,
\label{1}
\end{equation}
where $\mathcal{I}_m$ stands for electromagnetic field action. Furthermore, Lovelock gravity's action $\mathcal{I}_L$ is defined as 
\begin{equation}
\mathcal{I}_L=\frac{1}{2}\int d^dx\sqrt{-g}\bigg[\sum_{p=0}^{s}\frac{\alpha_p}{2^p}\delta^{\mu_1...\mu_{2p}}_{\nu_1...\nu_{2p}} R^{\nu_1\nu_2}_{\mu_1\mu_2}...R^{\nu_{2p-1}\nu_{2p}}_{\mu_{2p-1}\mu_{2p}}\bigg],
\label{2}
\end{equation}
where $R^{\alpha\beta}_{\mu\nu}$ are the components of curvature tensor, $\delta^{\mu_1...\mu_{2p}}_{\nu_1...\nu_{2p}}$ is the generalized Kronecker delta with order $2p$ and $s=[\frac{d-1}{2}]$ defines the maximum integer when the bracket represent the integer part. Furthermore, $\alpha_p$'s are arbitrary coupling parameters in which $\alpha_0$ is proportional to the cosmological constant $\Lambda$, i.e., $\alpha_0=-2\Lambda$.

Using variational principle, one can get the equations of motion as 
\begin{equation}
\sum_{p=0}^{s}\frac{\alpha_p}{2^{p+1}}\delta^{\nu\lambda_1...\lambda_{2p}}_{\mu\rho_1...\rho_{2p}} R^{\rho_1\rho_2}_{\lambda_1\lambda_2}...R^{\rho_{2p-1}\rho_{2p}}_{\lambda_{2p-1}\lambda_{2p}}=T^{\nu}_{\mu},
\label{4}
\end{equation}
in which $T^{\nu}_{\mu}$ refers to energy-momentum tensor and is defined as
\begin{equation}
T_{\mu\nu}=-\frac{2}{(-g)^{1/2}}\frac{\delta \mathcal{I}_m}{\delta g^{\mu\nu}}.
\label{5}
\end{equation}
 
 In order to determine the static solution of (\ref{4}), we are supposing the general spherically symmetric line element
\begin{equation}
ds^2=-f(r)dt^2+\frac{dr^2}{f(r)}+r^2 (h_{ab}dx^adx^b).
\label{6}
\end{equation}
Here $h_{ab}dx^adx^b$ is the metric of $(d-2)$-dimensional constant curvature space with a curvature $k=1,0,-1$, respectively, for spherical, flat and hyperbolic spaces.

The Lagrangian density characterizing double-logarithmic electrodynamics is defined as
 \begin{equation}\begin{split}
L_{m}&=\frac{1}{2\beta}\big[\big(1-\eta^{1/2}\big)\log{\big(1-\eta^{1/2}\big)}+\big(1+\eta^{1/2}\big)\log{\big(1+\eta^{1/2}\big)}\big].        \label{7}\end{split}
\end{equation}
 Here $\eta=-2\beta \mathcal{P}$ whereas, $\mathcal{P}=F_{\mu\nu}F^{\mu\nu}=2\big(\textbf{B}^2-\textbf{E}^2\big)$ such that $\textbf{E}$ is the electric field and $\textbf{B}$ denotes the magnetic field. By varying Eq. (\ref{1}) for gauge potential $A_{\mu}$, the field equations for NLED take the form
  \begin{equation}\begin{split}
 \nabla^{\rho}\bigg[\frac{\sqrt{-g}}{\eta^{1/2}}\log{\bigg(\frac{1-\eta^{1/2}}{1+\eta^{1/2}}\bigg)}F_{\rho\sigma}\bigg]=0.  \label{8}\end{split}
 \end{equation}
 
  The energy-momentum tensor associated with (\ref{7}) can be expressed as
 \begin{equation}\begin{split}
 T_{\rho\sigma}&=\frac{1}{2\beta}\bigg[\big(1-\eta^{1/2}\big)\log{\big(1-\eta^{1/2}\big)}+\big(1+\eta^{1/2}\big)\\&\times\log{\big(1+\eta^{1/2}\big)}\bigg]g_{\rho\sigma}-\frac{2F_{\rho\lambda}F^{\lambda}_{\sigma}}{\eta^{1/2}}\log{\bigg(\frac{1-\eta^{1/2}}{1+\eta^{1/2}}\bigg)}.        \label{9}\end{split}
 \end{equation}
 To derive the black hole solution with a pure magnetic charge, it is natural to take $\textbf{E}=0$ and $\textbf{B}\neq0$. This yields the Maxwell invariant equal to $\mathcal{P}=\frac{2Q^2}{r^{2d-4}}$. Now, let's suppose the coefficient $f(r)$ in the metric ansatz (\ref{6}) as
   \begin{equation}
   f(r)=k-r^2\Psi(r),\label{10}
   \end{equation}
   where $\Psi(r)$ can be find from the polynomial 
   \begin{equation}
   P[\Psi(r)]=\sum_{p=0}^{s}\overline{\alpha}_p\Psi^p(r).\label{11}
   \end{equation}
  Note that, the coefficients $\overline{\alpha}_p$'s are related to the coupling parameters $\alpha_p$'s through the equations
  \begin{align}\begin{split}
  &\overline{\alpha}_0=\frac{\alpha_0}{(d-1)(d-2)}, \\&
  \overline{\alpha}_1=1,\\&\overline{\alpha}_p=\prod_{i=3}^{2p}(d-i)\alpha_p,\label{12}\end{split}
  \end{align} 
 where the general value of $\overline{\alpha}_p$ in the above definition holds only for $p>1$. Therefore, by solving the field equations (\ref{4}) with an assumption, i.e., $\textbf{E}=0$ and $\textbf{B}\neq0$ in Eq. (\ref{9}), the metric function $f(r)$ will satisfy the polynomial equation  
    \begin{eqnarray}\begin{split}
    P[\Psi(r)]&=\frac{\mu}{r^{d-1}}-\frac{1}{\beta (d^2-3d+2)}\log{\bigg(1+\frac{4\beta Q^2}{r^{2(d-2)}}\bigg)}-\frac{8Q^2d}{(d^2-4d+3)r^{2(d-2)}}\\&F_1\bigg[1,\frac{d-3}{2(d-2)},\frac{3d-7}{2(d-2)},-\frac{4\beta Q^2}{r^{2(d-2)}}\bigg]+\frac{4Q^2 r^{2-d}}{\sqrt{\beta}(d^2-5d+6)}\tan^{-1}{\bigg(\frac{2Q\sqrt{\beta}}{r^{d-2}}\bigg)}. \label{13}\end{split}
    \end{eqnarray}
    Here $F_1$ stands for the Gaussian hypergeometric function and the integration constant $\mu$ is related to the finite black hole's mass through the formula \cite{138f}
     \begin{equation}
    M=\frac{(d-2)\mu\Sigma_{d-2}}{2},\label{13a}
    \end{equation}
     in which $\Sigma_{d-2}$ for the case $k=1$ is given by
    \begin{equation}
    \Sigma_{d-2}=\frac{2\pi^{\frac{(d-1)}{2}}}{\Gamma{\big(\frac{d-1}{2}\big)}}.\label{14}
    \end{equation}
    The radius of the outer horizon can be found from the condition $f(r_h)=0$. Hence, it is simple to obtain from Eq. (\ref{10}) that
    \begin{equation}
    r_h^2=\frac{k}{\Psi(r_h)}.\label{15}
    \end{equation}
    This yields the total mass as a function of $r_h$ in the following form
    \begin{eqnarray}\begin{split}
    M&=\frac{\Sigma_{d-2}}{2}\bigg[\sum_{p=0}^{s}\frac{\overline{\alpha}_pk^p(d-2)}{r_h^{-(d-2p-1)}}+\frac{8Q^2d(d-2)}{(d^2-4d+3)r_h^{d-3}}F_1\bigg[1,\frac{d-3}{2(d-2)},\frac{3d-7}{2(d-2)},-\frac{4\beta Q^2}{r_h^{2(d-2)}}\bigg]\\&+\frac{r_h^{d-1}}{\beta(d-1)}\log{\bigg(1+\frac{4\beta Q^2}{r_h^{2d-4}}\bigg)}-\frac{4Q^2r_h}{\sqrt{\beta}(d-3)}\tan^{-1}{\bigg(\frac{2Q\sqrt{\beta}}{r_h^{d-2}}\bigg)}\bigg]. \label{16}\end{split}
    \end{eqnarray}
    
     The Hawking temperature \cite{138} corresponding to the polynomial equation (\ref{13}) can be defined as $T_H=\kappa_s/2\pi$ in which $\kappa_s$ represents the surface gravity. Thus, one can get
     \begin{eqnarray}\begin{split}
   T_{H}&=\frac{1}{4\pi X(r_h)}\bigg[\sum_{p=0}^{s}\frac{\overline{\alpha}_pk^p(d-2p-1)}{r_h^{2p+1}}+\frac{1}{(d-2)r_h\beta}\log{\bigg(1+\frac{4\beta Q^2}{r_h^{2d-4}}\bigg)}\\&-\frac{8Q^2\big(Q(d-1)-d^2+2d-3\big)}{(d^2-4d+3)r_h(r_h^{2(d-2)}+4\beta Q^2)}+\frac{4Q^2\tan^{-1}{\bigg(\frac{2Q\sqrt{\beta}}{r_h^{d-2}}\bigg)}}{\sqrt{\beta}(d-2)(d-3)r_h^{d-1}}\bigg], \label{17}\end{split} 
    \end{eqnarray}
    where $X(r_h)$ is defined by
    \begin{equation}
    X(r_h)=\sum_{p=0}^{s}\frac{p\overline{\alpha}_pk^{p-1}}{r_h^{2p}}.\label{18}
    \end{equation}
    
     By using Wald's method \cite{139,140}, the entropy may be defined as
    \begin{equation}
    S=-2\pi\oint d^{d-2}x\sqrt{\gamma}\frac{\partial L}{\partial R_{abcd}}\epsilon_{ab}\epsilon{cd}. \label{19}
    \end{equation}
    Here, $L$ stands for the total Lagrangian density, $\epsilon_{ab}$ is the binormal to the horizon and the components $\gamma_{ij}$ define the metric tensor of the horizon. Thus, associated to the polynomial equation (\ref{13}), one may compute 
    \begin{equation}
    S=2\pi(d-2)\Sigma_{d-2}\sum_{p=1}^{s}\frac{\overline{\alpha}_ppk^{p-1}r_h^{d-2p}}{(d-2p)}.\label{20}
    \end{equation}
    
    The general formula for the description of heat capacity can be written as
    \begin{equation}
    C_Q=T_{H}(r_h)\frac{dS}{dT_H}|_{Q}. \label{21}
    \end{equation}
   Using Eq. (\ref{17}), it is straightforward to calculate  
     \begin{equation}\begin{split}
    \frac{\partial T_H}{\partial r_h}&=\frac{1}{4\pi X(r_h)}\bigg(\mathcal{A}'(r_h)-\sum_{p=0}^{s}\frac{\overline{\alpha}_pk^{p}(d-2p-1)(2p+1)}{r_h^{2p+2}}\bigg)\\&-\frac{X'(r_h)}{4\pi X^2(r_h)}\bigg(\sum_{p=0}^{s}\frac{\overline{\alpha_p}k^p(d-2p-1)}{r_h^{2p+1}}+\mathcal{A}(r_h)\bigg), \label{22} \end{split}
    \end{equation}
    where
    \begin{equation}\begin{split}
    X'(r_h)=-\sum_{p=1}^{s}\frac{2p^2\overline{\alpha_p}k^{p-1}}{r_h^{2p+1}}, \label{23},\end{split}
    \end{equation} 
    \begin{equation}\begin{split}
    \mathcal{A}(r_h)&=\frac{1}{\beta (d-2)r_h}\log{\bigg(1+\frac{4\beta Q^2}{r_h^{2d-4}}\bigg)}+\frac{128Q^2\big(Qd-Q-d^2+2d-3\big)}{(d^2-4d+3)(r_h^{2d-3}+4\beta Q^2r_h)}\\&-\frac{4Q^2}{\beta^{1/2}(d^2-5d+6)r_h^{d-1}}\arctan{\bigg(\frac{2\sqrt{\beta}Q}{r_h^{d-2}}\bigg)}. \label{24} \end{split}
    \end{equation}
    and
    \begin{equation}\begin{split}
    \mathcal{A}'(r_h)&=\frac{16 Q^2}{(d^2-4d+1)(r_h^{2d-3}+4\beta Q^2r_h)^2}\bigg(\big(d^3-4d^2-Qd(d-3)+\\&2d+(3-2Q)\big) r_h^{2(d-2)}+(4d-12)\beta Q^2\bigg)+\frac{4 (d-1)Q^2}{\sqrt{\beta}(d^2-5d+6)r_h^{d}}\\&\times\arctan{\bigg(\frac{2Q\beta^{1/2}}{r_h^{d-2}}\bigg)}-\frac{1}{\beta r_h^2(d-2)}\log{\bigg(1+\frac{4\beta Q^2}{r_h^{2d-4}}\bigg)}. \label{25} \end{split}
    \end{equation}
    By using Eqs. (\ref{20})-(\ref{25}) one may get
   \begin{eqnarray}\begin{split}
   C_Q&=\frac{4\pi(d-2)\Sigma_{d-2}W(r_h)\sum_{p=1}^{n-1}pk^{p-1}\overline{\alpha}_pr_h^{d-2p-1}\big(\mathcal{A}(r_h)+\sum_{p=0}^{s}\frac{\overline{\alpha}_pk^p(d-2p-1)}{r_h^{2p+1}}\big)}{W(r_h)\Xi(r_h)-W'(r_h)\big(\mathcal{A}(r_h)+\sum_{p=0}^{s}\frac{\overline{\alpha}_pk^p}{r_h^{2p+1}}\big)}, \label{26}\end{split}
   \end{eqnarray}
   where
   \begin{equation}
   \Xi(r_h)=\mathcal{A}'(r_h)-\sum_{p=0}^{s}\frac{\overline{\alpha}_pk^p(2p+1)(d-2p-1)}{r_h^{2p+2}}.\label{26a}
   \end{equation}
  
  The heat capacity is very significant as it exposes local thermodynamic stability in the canonical ensemble. In other words, black hole will be stable if $C_Q>0$ and unstable when $C_Q<0$.

\section{Einsteinian black holes and double-logarithmic electrodynamics} 

 Here I wish to discuss the higher dimensional black holes of Einstein's theory. To do this, the higher curvature terms should be eliminated from the action (\ref{1}), i.e., by choosing $\alpha_p=0$ for all values of $p\geq 2$. Note that, the gravitational field is once more coupled to NLED. Hence, one can get the solution from (\ref{13}) as
      \begin{eqnarray}\begin{split}
      f(r)&=1-\frac{\mu}{r^{d-3}}+\frac{\alpha_0r^2}{d^2-3d+2}+\frac{8dQ^2}{(d^2-4d+3)r^{2d-6}}\\&\times F_1\bigg[1,\frac{d-3}{2(d-2)},\frac{3d-7}{2(d-2)},\frac{-4\beta Q^2}{r^{2(d-2)}}\bigg]+\frac{r^2}{\beta (d^2-3d+2)}\log{\bigg(1+\frac{4\beta Q^2}{r^{2d-4}}\bigg)}\\&-\frac{4Q}{\beta^{1/2}(d^2-5d+6)r^{d-4}}\arctan{\bigg(\frac{2Q\beta^{1/2}}{r^{d-2}}\bigg)}. \label{27}\end{split}
      \end{eqnarray}
 
  For $d=4$, this becomes
      \begin{eqnarray}\begin{split}
      f(r)&=1-\frac{\mu}{r}-\frac{32Q^{2}}{3r^{2}}-\frac{\Lambda r^2}{3}-\frac{4 Q^{\frac{3}{2}}}{3r\beta^{\frac{1}{4}}}\bigg[\arctan{\bigg(1+\frac{r}{\sqrt{Q}\beta^{\frac{1}{4}}}\bigg)}\\&-\arctan{\bigg(1-\frac{r}{\sqrt{Q}\beta^{\frac{1}{4}}}\bigg)}\bigg]+\frac{r^2}{6\beta}\log{\bigg(1+\frac{4\beta Q^2}{r^4}\bigg)}-\frac{2Q}{\sqrt{\beta}}\\&\times\arctan{\bigg(\frac{2Q\sqrt{\beta}}{r^2}\bigg)}-\frac{2Q^{\frac{3}{2}}}{3\beta^{\frac{1}{4}}r}\log{\bigg[\frac{r^2-2\sqrt{Q}r\beta^{\frac{1}{4}}+2Q\sqrt{\beta}}{r^2+2\sqrt{Q}r\beta^{\frac{1}{4}}+2Q\sqrt{\beta}}\bigg]},\label{28}\end{split}
      \end{eqnarray}  
which describes the non-asymptotically flat black holes of four-dimensional Einstein's gravity with double-logarithmic electromagnetic source. In the limit $\beta\rightarrow 0$, the solution (\ref{27}) reduces to the magnetized black hole solution of Einstein-Maxwell theory, i.e.,
\begin{eqnarray}\begin{split}
f(r)&=1-\frac{\mu}{r^{d-3}}-\frac{2\Lambda r^2}{d^2-3d+2}+\frac{8Q^{2}d}{(d^2-4d+3)r^{2d-6}}+O(\beta).\label{29}\end{split}
\end{eqnarray}  

For the metric ansatz (\ref{6}), the Kretschmann and Ricci scalars can be obtained as

\begin{eqnarray}\begin{split}
R&=\bigg((d^2-5d+6)\bigg(\frac{1-f(r)}{r^2}\bigg)-f''(r)-\frac{(2d-4)}{r}f'(r)\bigg),\label{30}\end{split}
\end{eqnarray}
and
\begin{eqnarray}\begin{split}
K&=\bigg(2(d^2-5d+6)\bigg(\frac{1-f(r)}{r^2}\bigg)^2-f''(r)^2+\frac{(2d-4)}{r^2}f''(r)^2\bigg),\label{31}\end{split}
\end{eqnarray}
where primes denote differentiation with respect to coordinate $r$. One can easily verify from (\ref{27}) that both of the above expressions diverge at $r=0$. Hence, it is concluded that $r=0$ is a true curvature singularity for solution (\ref{27}).
 
Using Eq. (\ref{13a}), one can find the mass as 
    \begin{eqnarray}\begin{split}
     M&=\frac{\Sigma_{d-2}}{2}\bigg[(d-2)r_h^{d-3}-\frac{2\Lambda r_h^{d-1}}{(d-1)}+\frac{8Q^2d(d-2)}{(d-1)(d-3)r_h^{d-3}}F_1\bigg[1,\frac{d-3}{2d-4},\frac{3d-7}{2d-4},-\frac{4\beta Q^2}{r_h^{2d-4}}\bigg]\\&+\frac{r_h^{d-1}}{\beta (d-1)}\log{\bigg(1+\frac{4\beta Q^2}{r_h^{2d-4}}\bigg)}-\frac{4Qr_h}{\sqrt{\beta}(d-3)}\arctan{\bigg(\frac{2Q\sqrt{\beta}}{r_h^{d-2}}\bigg)}\bigg].\label{32}\end{split}
          \end{eqnarray}
           \begin{figure}[h]
          	\centering
          	\includegraphics[width=0.8\textwidth]{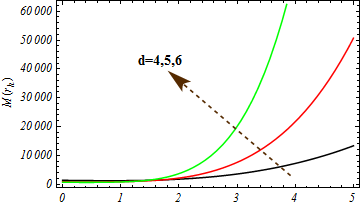}
          	\caption{ The plots show how the mass $M$ (Eq. (\ref{32})) varies with horizon radius $r_h$. The parameters are selected as $Q=1.5$, $\beta=1.7$, $\Sigma_{d-2}=100$ and $\Lambda=-3$.}\label{skr1}
          \end{figure} 
          \begin{figure}[h]
          	\centering
          	\includegraphics[width=0.8\textwidth]{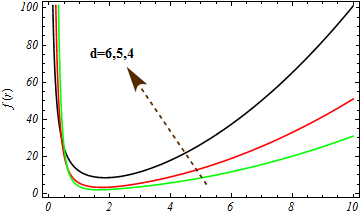}
          	\caption{The plots showing the behavior of $f(r)$ (Eq. (\ref{27})) in different dimensions. Particular values of the parameters are considered as $m=10$, $Q=1.50$, $\beta=1.7$, $\Sigma_{d-2}=100$ and $\Lambda=-3$.} \label{Askar1a}
          \end{figure} 
          \begin{figure}[h]
          	\centering
          	\includegraphics[width=0.8\textwidth]{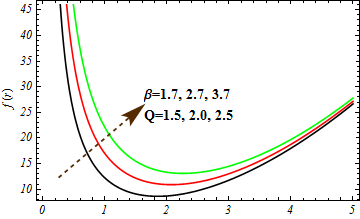}
          	\caption{The plots showing the behavior of $f(r)$ (Eq. (\ref{27})) with various choices of $\beta$ and $Q$. Particular values of parameters are considered as $m=1$, $d=6$, $\Sigma_{d-2}=100$ and $\Lambda=-3$.} \label{par1b}
          \end{figure} 
          
     The mass for different spacetime dimensions is plotted in Fig. \ref{skr1}. Those values of $r_h$ which correspond to the positivity of mass function imply the black holes of such horizon radii can exist. The solution (\ref{27}) for fixed values of parameters $Q$ and $\beta$ in different dimensional geometries is plotted in Fig. \ref{Askar1a}. Similarly, the effects of charge and nonlinear magnetic field on the metric function (\ref{27}) can be seen from Fig. \ref{par1b}.
   One could use Eq. (\ref{27}) to determine the Hawking temperature $T_H(r_h)=\kappa_s/2\pi$ as 
    \begin{eqnarray}\begin{split}
    T_H(r_h)&=\frac{1}{4\pi}\bigg[\frac{d-3}{r_h}-\frac{2\Lambda r_h(d-1)}{(d^2-3d+2)}+\frac{r_h}{\beta (d-2)}\log{\bigg(1+\frac{4\beta Q^2}{r_h^{2d-4}}\bigg)}\\&-\frac{4Qr_h^{3-d}}{(d^2-5d+6)\sqrt{\beta}}\arctan{\bigg(\frac{2Q\sqrt{\beta}}{r_h^{d-2}}\bigg)}-\frac{8Q^2r_h(d(d-3)-2)}{(d^2-4d+3)(r_h^{2(d-2)}+4\beta Q^2)}\bigg]. \label{33}\end{split}
    \end{eqnarray} 
    \begin{figure}[h]
    	\centering
    	\includegraphics[width=0.8\textwidth]{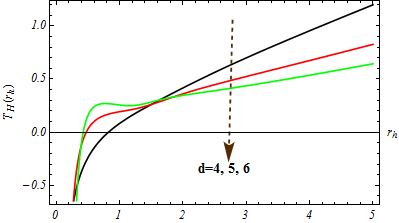}
    	\caption{The plots describe the changes in temperature $T_H$ (Eq. (\ref{33})) in different spacetime dimensions. The parameters are taken as $Q=1.5$, $\beta=1.7$, $\Sigma_{d-2}=100$ and $\Lambda=-3$.}\label{Saif1a}
    \end{figure} 
\begin{figure}[h]
	\centering
	\includegraphics[width=0.8\textwidth]{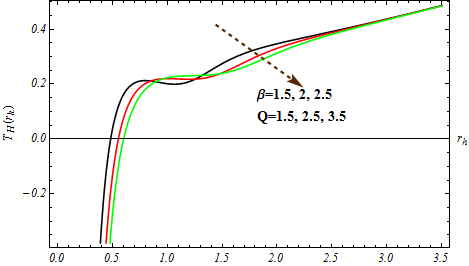}
	\caption{The plots describe the changes in temperature $T_H$ (Eq. (\ref{33})) with different choices of $Q$ and $\beta$. The other parameters are taken as $d=6$, $\Sigma_{d-2}=100$ and $\Lambda=-3$.}\label{mma1}
\end{figure}

The Hawking temperature for different values of $d$ and fixed values of other parameters is plotted in Fig. \ref{Saif1a}. Those values of $r_h$ which correspond to non-negative temperature indicate the black holes with such horizons are physical. In addition, Fig. \ref{mma1} demonstrates how nonlinearity and charge parameters affect the black hole's temperature. It is clear that these parameters inevitably affect the temperature of smaller black holes. However, these effects could be ignored for the black holes of larger size. The expression of entropy for the solution (\ref{27}) can be computed from $S=\int T_H^{-1}dM$ as
\begin{equation}
S=2\pi\Sigma_{d-2}r_h^{d-2},\label{34}
\end{equation}
which clearly shows that Hawking area law holds for the obtained magnetized Einsteinian black holes.

Differentiation of Hawking temperature (\ref{33}) yields
     \begin{eqnarray}\begin{split}
     \frac{dT_H}{dr_h}&=\frac{1}{4\pi}\bigg[-\frac{2\Lambda}{(d-2)}-\frac{d-3}{r_h^2}+\frac{1}{\beta (d-2)}\log{\bigg(1+\frac{4\beta Q^2}{r_h^{2d-4}}\bigg)}\\&+\frac{4Qr_h^{2-d}}{(d-2)}\arctan{\bigg(\frac{2Q\sqrt{\beta}}{r_h^{d-2}}\bigg)}+\frac{16Q^2r_h^{2(d-2)}(d-2)(d(d-3)-2)}{(d-1)(d-3)(r_h^{2(d-2)}+4\beta Q^2)^2}\\&-\frac{8Q^2(2d^2-8d+4)}{(d-1)(d-3)(r_h^{2d-4}+4\beta Q^2)}\bigg].\label{35}\end{split}
     \end{eqnarray}
     
      The heat capacity can be determined by utilizing Eqs. (\ref{33})-(\ref{35}) in Eq. (\ref{21}) as
     \begin{eqnarray}\begin{split}
    C_Q&=\frac{2\pi (d-2)\Sigma_{d-2}r_h^{d-2}\bigg((d^2-5d+6)-2\Lambda r_h^2+(d-2)r_h\zeta_1\bigg)}{-2\Lambda r_h^2-(d-3)(d-2)+(d-2)r_h^2\zeta_2}, \label{36}\end{split}
     \end{eqnarray}
     where
     \begin{eqnarray}\begin{split}
     \zeta_1(r_h)&=\frac{r_h}{\beta(d-2)}\log{\bigg(1+\frac{4\beta Q^2}{r_h^{2d-4}}\bigg)}-\frac{4Q}{(d^2-5d+6)\sqrt{\beta}r_h^{d-3}}\arctan{\bigg(\frac{2Q\sqrt{\beta}}{r_h^{d-2}}\bigg)}\\&-\frac{8Q^2(d^2-3d-2)r_h}{(d-1)(d-3)(r_h^{2d-4}+4\beta Q^2)}, \label{37}\end{split}
     \end{eqnarray}
  and
  \begin{eqnarray}\begin{split}
  \zeta_2(r_h)&=\frac{1}{\beta(d-2)}\log{\bigg(1+\frac{4\beta Q^2}{r_h^{2d-4}}\bigg)}+\frac{4Q}{(d-2)r_h^{d-2}}\arctan{\bigg(\frac{2Q\sqrt{\beta}}{r_h^{d-2}}\bigg)}\\&-\frac{8Q^2(2d^2-8d+4)}{(d-1)(d-3)(r_h^{2d-4}+4\beta Q^2)}+\frac{8Q^2(2d-4)(d^2-3d-2)r_h^{2d-4}}{(d^2-4d+3)(r_h^{2d-4}+4\beta Q^2)^2}. \label{38}\end{split}
  \end{eqnarray}
  \begin{figure}[h]
  	\centering
  	\includegraphics[width=0.8\textwidth]{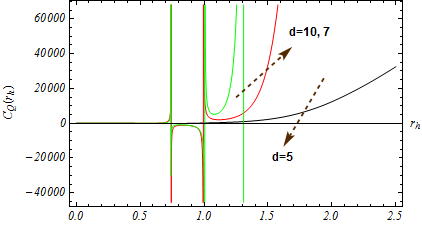}
  	\caption{The plots describe the changes of specific heat $C_Q$ (Eq. (\ref{36})) in different spacetime dimensions. The parameters are taken as $Q=1.5$, $\beta=0.5$, $\Sigma_{d-2}=100$ and $\Lambda=-3$.}\label{khan1}
  \end{figure}
\begin{figure}[h]
	\centering
	\includegraphics[width=0.8\textwidth]{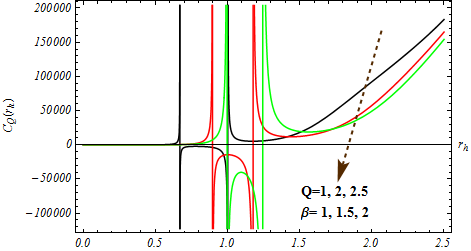}
	\caption{ The plots describe the effects of $\beta$ and $Q$ on specific heat $C_Q$ (Eq. (\ref{36})). The other parameters are chosen as $d=6$, $\Sigma_{d-2}=100$ and $\Lambda=-3$.}\label{khan1q}
\end{figure}
\begin{figure}[h]
\centering
\includegraphics[width=0.8\textwidth]{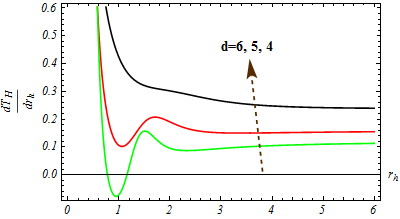}
\caption{The plots describe the variations of $\frac{dT_H}{dr_h}$ (Eq. (\ref{35})) in different dimensions. The particular values of parameters are considered as $Q=1.5$, $\beta=1.7$, $\Sigma_{d-2}=100$ and $\Lambda=-3$.}\label{change1}
\end{figure}
\begin{figure}[h]
	\centering
	\includegraphics[width=0.8\textwidth]{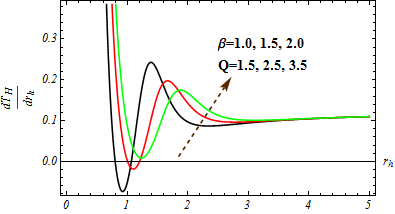}
	\caption{The plots describe the effects of $\beta$ and $Q$ on the quantity $\frac{dT_H}{dr_h}$ (Eq. (\ref{35})). The other parameters are fixed as $d=6$, $\Sigma_{d-2}=100$ and $\Lambda=-3$.}\label{change1a}
\end{figure}

 The variations of heat capacity with respect to the outer horizon with different $d$ are plotted in Fig. \ref{khan1}. Similarly, its dependence on the parameters $\beta$ and $Q$ can be analyzed from Fig. \ref{khan1q}. The region with $C_Q>0$ illustrates the local thermodynamic stability in the canonical ensemble. The point at which it vanishes specifies the first-order phase transition, and the region in which it is smaller than zero demonstrates the instability of black holes. However, those values of $r_+$ at which it is singular give second-order phase transition points. These singular points of $C_Q$ can also be described from the roots of $\frac{dT_H}{dr_h}=0$. The behavior of $dT_H/dr_h$ in different dimensional spacetimes is demonstrated in Fig. \ref{change1}. The points of intersection between the curve of $\frac{dT_H}{dr_h}$ and $r_h$-axis exhibit the second-order phase transition points. Furthermore, the impacts of the nonlinearity and magnetic charge on the divergences of $C_Q$ can be seen in Fig. \ref{change1a}.

\section{Gauss-Bonnet black holes and double-logarithmic electrodynamics}

   In this section, the nonlinearly charged Gauss-Bonnet black holes and their thermodynamics are covered. To derive the metric function for these black holes, it is important to take $\alpha_2\neq 0$ and $\alpha_p=0$ for $p\geq 3$ in Eq. (\ref{1}). From Eq. (\ref{13}), this yields the solution in two branches
  \begin{eqnarray}\begin{split}
  f_{\pm}(r)&=k+\frac{r^2}{2\overline{\alpha}_2}\bigg(1\pm\sqrt{H(r)}\bigg), \label{39}\end{split}
  \end{eqnarray}
  where $H(r)$ is obtained as
  \begin{eqnarray}\begin{split}
  H(r)&=1+\frac{8\Lambda\overline{\alpha}_2}{(d^2-3d+2)}+\frac{4\overline{\alpha}_2\mu}{r^{d-1}}-\frac{32\overline{\alpha}_2dQ^2}{(d^2-4d+3)r^{2d-4}}\\&\times F_1\bigg[1,\frac{d-3}{2(d-2)},\frac{3d-7}{2(d-2)},\frac{-4\beta Q^2}{r^{2(d-2)}}\bigg]-\frac{4\overline{\alpha}_2}{\beta(d^2-3d+2)}\log{\bigg(1+\frac{4\beta Q^2}{r^{2d-4}}\bigg)}\\&+\frac{16\overline{\alpha}_2Q}{\sqrt{\beta}(d^2-5d+6)r^{d-2}}\tan^{-1}{\bigg(\frac{2Q\sqrt{\beta}}{r^{d-2}}\bigg)}, \label{40}\end{split}
  \end{eqnarray}
  and $\overline{\alpha}_2=(d-3)(d-4)\alpha_2$.
  The asymptotic value of $f_{\pm}(r)$ when $r\rightarrow\infty$ can be given as
  \begin{eqnarray}\begin{split}
  f_{\pm}(r)&=1+\frac{r^2}{2\overline{\alpha}_2}\bigg(1\pm\sqrt{1+\frac{8\overline{\alpha}_2\Lambda}{(d-1)(d-2)}}\bigg), \label{41}\end{split}
  \end{eqnarray}
  which depending on $\Lambda$ it is dS, AdS or flat. In this case too, both expressions given in (\ref{30}) and (\ref{31}) are singular at $r=0$. Hence, we conclude that the solution (\ref{39}) describes the nonlinearly charged black hole with magnetic charge $Q$. The finite mass of the black hole can be work out from the zeros of $f(r_h)$ and Eq. (\ref{13a}) as follows
  
  \begin{eqnarray}\begin{split}
  M&= \frac{\Sigma_{d-2}}{2}\bigg[\alpha_2(d^2-5d+6)(d-4)r_h^{d-5}+r_h^{d-3}-\frac{2\Lambda r_h^{d-1}}{(d-1)}+\frac{r_h^{d-1}}{\beta (d^2-3d+2)}\\&\log{\bigg(1+\frac{4\beta Q^2}{r_h^{2d-4}}\bigg)}+\frac{8Q^2d}{(d^2-4d+3)r_h^{d-3}}F_1\bigg[1,\frac{d-3}{2(d-2)},\frac{3d-7}{2(d-2)},\frac{-4\beta Q^2}{r_h^{2(d-2)}}\bigg]\\&-\frac{4Qr_h}{\sqrt{\beta}(d-3)(d-4)}\tan^{-1}{\bigg(\frac{2Q\sqrt{\beta}}{r_h^{d-2}}\bigg)}\bigg].\label{42}\end{split}
  \end{eqnarray} 
  \begin{figure}[h]
  	\centering
  	\includegraphics[width=0.8\textwidth]{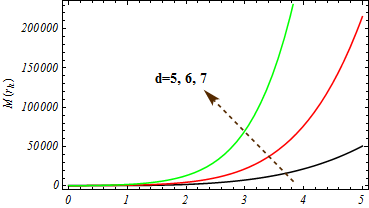}
  	\caption{The plots show how the mass $M$ (Eq. (\ref{42})) varies with horizon radius $r_h$. The parameters are considered as $Q=0.5$, $\beta=0.3$, $\Sigma_{d-2}=100$, $\Lambda=-3$ and $\alpha_2=0.5$.}\label{MGB1}
  \end{figure} 
\begin{figure}[h]
	\centering
	\includegraphics[width=0.8\textwidth]{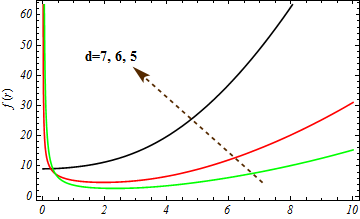}
	\caption{The plots showing the behavior of $f_+(r)$ (Eq. (\ref{39})) in different dimensions. Particular values of the parameters are considered as $m=10^4$, $Q=0.3$, $\beta=0.3$, $k=1$, $\Sigma_{d-2}=100$, $\Lambda=-3$ and $\alpha_2=0.5$.}\label{Askar2d}
\end{figure}
\begin{figure}[h]
	\centering
	\includegraphics[width=0.8\textwidth]{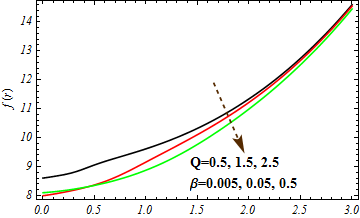}
	\caption{The plots showing the behavior of $f_+(r)$ (Eq. (\ref{39})) with different values of $\beta$ and $Q$. Particular values of parameters are selected as $m=10^4$, $d=5$, $\Sigma_{d-2}=100$, $k=1$, $\Lambda=-3$ and $\alpha_2=0.5$.}\label{Askar2Q}
\end{figure}  
  Figs. \ref{MGB1} shows the behaviors of the total mass with different values of $d$. The positions of the horizon radii can be thought of as those values of $r_h$ that give the non-negative mass. Figs. \ref{Askar2d} and \ref{Askar2Q} show how the solution (\ref{39}) behaves as a function of $r$. It is shown that the black holes are impacted by the parameters $\beta$, $Q$, and $d$ in an inevitable manner. The radius of the event horizon is the value where the curve associated with (\ref{39}) meets with the $r$-axis for fixed values of parameters $\beta$, $Q$, and $d$. 
  
 The expression of Hawking temperature $T_H$ can be work out as
  \begin{eqnarray}\begin{split}
  4\pi T_H(r_h)&=-\frac{2}{r_h}+\frac{r_h^4(r_h^2+2\alpha_2(d-3)(d-4)^{-1}}{4\alpha_2(d-3)(d-4)}\bigg[\frac{4\alpha_2^2(d-3)^2(d-4)^2(d-1)}{r_h^5}\\&+\frac{4(d^2-7d+12)\alpha_0\alpha_2}{(d-2)r_h}+\frac{4\alpha_2(d^2-7d+12)}{\beta (d-2)r_h}\log{\bigg(1+\frac{4\beta Q^2}{r_h^{2d-4}}\bigg)}\\&-\frac{16Q^2\alpha_2(d-4)}{\sqrt{\beta}(d-2)r_h^{d-1}}\arctan{\bigg(\frac{2Q\sqrt{\beta}}{r_h^{d-2}}\bigg)}-\frac{32(d-4)(d^2-3d-2)Q^2\alpha_2}{(d-1)(r_h^{2d-3}+4\beta Q^2r_h)}\bigg].\label{43}\end{split}
  \end{eqnarray} 
    \begin{figure}[h]
	\centering
	\includegraphics[width=0.8\textwidth]{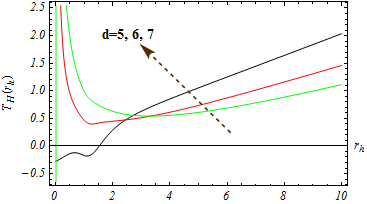}
	\caption{The plots describe the changes in temperature $T_H$ (Eq. (\ref{43})) in different spacetime dimensions. The parameters are taken as $Q=1.5$, $\beta=0.5$, $\Sigma_{d-2}=100$, $\Lambda=-3$ and $\alpha_2=2$.}\label{Saif2a}
\end{figure} 
\begin{figure}[h]
	\centering
	\includegraphics[width=0.8\textwidth]{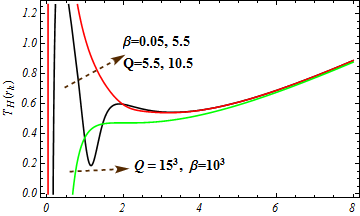}
	\caption{The plots describe the changes in temperature $T_H$ (Eq. (\ref{43})) with different values of $Q$ and $\beta$. The other parameters are taken as $d=7$, $\Sigma_{d-2}=100$, $\Lambda=-3$ and $\alpha_2=2$.}\label{Saif2aq}
\end{figure} 
 
  Fig. \ref{Saif2a} demonstrates the behavior of Hawking temperature in various dimensions. The black hole with horizon radius $r_h$ must be physical if the value of $r_h$ that results in $T_H\geq0$ is true. The impacts of $\beta$ and $Q$ on the black hole's temperature can be observed from Fig. \ref{Saif2aq}. It is obvious that as the black hole's size grows, these effects become smaller and eventually nonexistent. According to Wald's approach \cite{139,140}, the entropy of the black hole is
  \begin{eqnarray}
  S=(2d-4)\pi\Sigma_{d-2}r_h^{d-4}\bigg(\frac{2r_h^2}{d-2}+2\alpha_2(d-3)\bigg).\label{44}
  \end{eqnarray}
  It is simple to calculate
  \begin{eqnarray}\begin{split}
  4\pi\frac{dT_H}{dr_+}&=\frac{2}{r_h^2}+\frac{(d-4)r_h^4Z_1(r_h)}{r_h^2+2(d-4)(d-3)\alpha_2}-\frac{132(d-4)^2(d-3)Q\alpha_2r_h^{6-d}}{\sqrt{\beta}(d-2)(r_h^2+2(d-4)(d-3)\alpha_2)^2}\\&\times \arctan{\bigg(\frac{2\sqrt{\beta}Q}{r_h^{d-2}}\bigg)}+\frac{8(d-4)r_h^5(3r_h^2+4(d-3)(d-4)\alpha_2)Z_2(r_h)}{(r_h^2+2(d^2-7d+12)\alpha_2)^2},\label{45}\end{split}
  \end{eqnarray}
  where
  \begin{eqnarray}\begin{split}
  Z_2(r_h)&=\frac{8(d^2-3d-2)Q^2}{(d-1)r_h(r_h^{2d-4}+4Q^2\beta)}+\frac{(d-4)(d^2-6d+9)(d-1)\alpha_2}{r_h^5}-\frac{(d-3)}{(d-2)r_h^3}\\&\times\big(-2\Lambda r_h^2+d^2-3d+2\big)-\frac{(d-3)}{(d-4)\beta r_h}\log{\bigg(1+\frac{4\beta Q^2}{r^{2d-4}}\bigg)},\label{46}\end{split}
  \end{eqnarray}
  \begin{eqnarray}\begin{split}
  Z_1(r_h)&=\frac{4(d-1)Q}{(d-2)\sqrt{\beta}r_h^6}\arctan{\bigg(\frac{2\sqrt{\beta}Q}{r_h^{d-2}}\bigg)}+\frac{16Q^2(1+d^3-5d^2+5d)}{(d-1)r_h^2(r_h^{2d-4}+4\beta Q^2)}\\&-\frac{64(d-2)(d^2-3d-2)Q^4\beta}{(d-1)r_h^2(r_h^{2d-4}+4\beta Q^2)^2}-\frac{(d-3)(-2\Lambda r_h^2+3d^2-9d+6)}{(d-2)r_h^4}\\&-\frac{5(d-4)(d-3)^2(d-1)\alpha_2}{r_h^6}-\frac{(d-3)}{(d-2)\beta r_h^2}\log{\bigg(1+\frac{4\beta Q^2}{r_h^{2d-4}}\bigg)}.\label{47}\end{split}
  \end{eqnarray}
 Hence, the heat capacity can be calculated from Eq. (\ref{21}) as follows:
 \begin{eqnarray}\begin{split}
 C_Q&=\frac{2\pi(d-2)\big(-8\alpha_2(d-3)(d-4)(r_h^2+2\alpha_2(d-3)(d-4))+r_h^5\Delta_1(r_h)\big)}{4r_h\alpha_2(d-4)(d-3)(r_h^2+2\alpha_2(d-3)(d-4))^{-1}}\\&\times\frac{\Sigma_{d-2}(r_h^{d-3}+2\alpha_2(d-3)(d-4)r_h^{d-5})}{\big((r_h^2+2\alpha_2(d-3)(d-4))^2\Delta_2(r_h)+8(d-4)r_h^5(3r_h^2+4(d-4)(d-3)\alpha_2)Z_2(r_h)\big)},\label{48}\end{split}
 \end{eqnarray}
 where 
 \begin{eqnarray}\begin{split}
 \Delta_1(r_h)&=\frac{4\alpha_2^2(d-3)^2(d-4)^2(d-1)}{r_h^5}-\frac{8(d-3)(d-4)\Lambda\alpha_2}{(d-2)r_h}\\&+\frac{4\alpha_2(d-3)(d-4)}{\beta (d-2)r_h}\log{\bigg(1+\frac{4\beta Q^2}{r_h^{2d-4}}\bigg)}-\frac{16Q\alpha_2(d-4)}{\sqrt{\beta}(d-2)r_h^{d-1}}\\&\times\arctan{\bigg(\frac{2Q\sqrt{\beta}}{r_h^{d-2}}\bigg)}-\frac{32(d-4)(d^2-3d-2)Q^2\alpha_2}{(d-1)(r_h^{2d-3}+4\beta Q^2r_h)},\label{49}\end{split}
 \end{eqnarray}
 and 
 \begin{eqnarray}\begin{split}
 \Delta_2(r_h)&=\frac{(d-4)r_h^4Z_1(r_h)}{(r_h^2+2\alpha_2(d-3)(d-4))}-\frac{132Q\alpha_2(d-4)^2(d-3)r_h^{6-d}}{\sqrt{\beta}(d-2)(r_h^2+2\alpha_2(d-3)(d-4))^2}\\&\times\arctan{\bigg(\frac{2Q\sqrt{\beta}}{r_h^{d-2}}\bigg)}+\frac{2}{r_h^2}.\label{50}\end{split}
 \end{eqnarray}
\begin{figure}[h]
	\centering
	\includegraphics[width=0.8\textwidth]{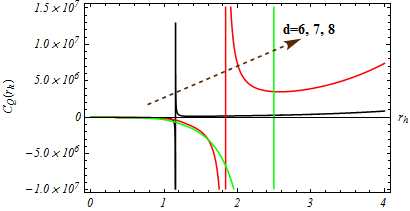}
	\caption{The plots describe the changes of specific heat $C_Q$ (Eq. (\ref{48})) in different spacetime dimensions. The parameters are taken as $Q=1.5$, $\beta=0.5$, $\Sigma_{d-2}=100$, $\Lambda=-3$ and $\overline{\alpha}_2=2(d-3)(d-4)$.}\label{khan2}
\end{figure}
\begin{figure}[h]
	\centering
	\includegraphics[width=0.8\textwidth]{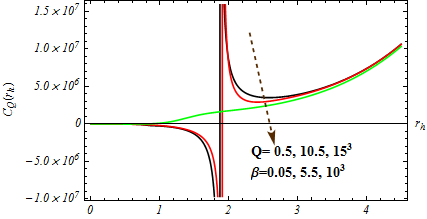}
	\caption{The plots describe the effects of $\beta$ and $Q$ on specific heat $C_Q$ (Eq. (\ref{48})). The other parameters are chosen as $d=7$, $\Sigma_{d-2}=100$, $\Lambda=-3$ and $\overline{\alpha}_2=2(d-3)(d-4)$.}\label{khan2q}
\end{figure}
\begin{figure}[h]
	\centering
	\includegraphics[width=0.8\textwidth]{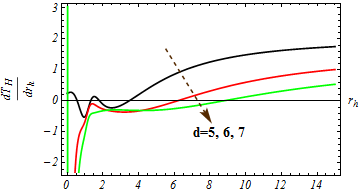}
	\caption{The plots describe the variations of $\frac{dT_H}{dr_h}$ (Eq. (\ref{46})) in different dimensions. The particular values of parameters are considered as $Q=1.5$, $\beta=0.5$, $\Sigma_{d-2}=100$, $\Lambda=-3$ and $\overline{\alpha}_2=2(d-3)(d-4)$.}\label{khan3}
\end{figure}
\begin{figure}[h]
	\centering
	\includegraphics[width=0.8\textwidth]{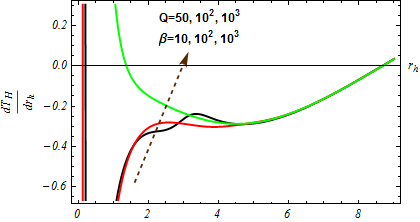}
	\caption{The plots describe the effects of $\beta$ and $Q$ on $\frac{dT_H}{dr_h}$ (Eq. (\ref{46})). The other parameters are considered as $d=7$, $\Sigma_{d-2}=100$, $\Lambda=-3$ and $\overline{\alpha}_2=2(d-3)(d-4)$.}\label{khan3q}
\end{figure}

 The region of local stability and the phase transition points can be identified from Figs. \ref{khan2} and \ref{khan2q}. The black hole with such a horizon radius is locally stable, as evidenced by the value of $r_h$ when $C_Q$ is non-negative. Moreover, the first-order and second-order phase transitions correspond to the zeros and singularities of this quantity. In addition to this, the second order phase transitions can also be understood from Figs. \ref{khan3} and \ref{khan3q} because the roots of an equation $dT_H/dr_h=0$ correspond to divergences of specific heat. Furthermore, one can conclude that the parameters $\beta$ and $Q$ both influence the local stability and phase transitions.
 
 \section{Third order Lovelock black holes and double-logarithmic electrodynamics}
 In this section, I want to study the third order Lovelock black holes supported by double-logarithmic electromagnetic field. To construct the solution corresponding to these black holes, it is convenient to assume $\alpha_3\neq 0$ and $\alpha_p=0$ for $p\geq 4$ in the calculated polynomial equation (\ref{13}). Note that for the sake of simplicity, we are also imposing the condition on coupling parameters as $\overline{\alpha}_3=\overline{\alpha}_2^2/3$. Hence, the metric function in this specific case can be obtained from Eq. (\ref{13}) as 
 \begin{equation}
 f(r)=1+\frac{r^2}{(d^2-7d+12)\alpha_2}\bigg(1-\sqrt[3]{\Theta(r)}\bigg), \label{51}
 \end{equation}
 where 
 \begin{eqnarray}\begin{split}
 \Theta(r)&=1+\Lambda\overline{\alpha}_2+\frac{3\mu\overline{\alpha}_2}{r^{d-1}}-\frac{24\overline{\alpha}_2dQ^2}{(d^2-4d+3)r^{2(d-2)}}\\&\times F_1\bigg[1,\frac{d-3}{2(d-2)},\frac{3d-7}{2(d-2)},\frac{-4\beta Q^2}{r^{2(d-2)}}\bigg]-\frac{3\overline{\alpha}_2}{\beta(d^2-3d+2)}\log{\bigg(1+\frac{4\beta Q^2}{r^{2d-4}}\bigg)}\\&+\frac{12\overline{\alpha}_2Q^2}{\beta^{1/2}(d^2-5d+6)r^{d-2}}\arctan{\bigg(\frac{2Q\sqrt{\beta}}{r^{d-2}}\bigg)}.
 \end{split}
 \label{52}
 \end{eqnarray}
 Note that, I have used $k=1$ in the above equation (\ref{51}). The asymptotic expression for this solution in the limit $r\rightarrow\infty$ can be obtained as
 \begin{eqnarray}\begin{split}
 f(r)&=1+\frac{r^2}{\overline{\alpha}_2}\bigg(1-\big(1+\overline{\alpha}_2\Lambda\big)^{1/3}\bigg), \label{53}\end{split}
 \end{eqnarray}
 which corresponds to the dS, AdS, or flat spacetimes according to the values of $\Lambda$. It is also worth noting that for small values of Lovelock parameter $\overline{\alpha}_2$, $f(r)$ tends to the value $(1-\Lambda/3r^2)$. Moreover, one can verify that both the expressions (\ref{30}) and (\ref{31}) are singular at the origin. Therefore, the resulting solution (\ref{51}) describes the metric function of magnetized third order Lovelock black hole. The mass $M$ dependent on the horizon radius $r_h$ can be once more calculated from the condition $f(r_h)=0$ and Eq. (\ref{13a}) as
 \begin{equation}\begin{split}
 M&=\frac{(d-2)\Sigma_{d-2}r^{d-1}}{6\overline{\alpha}_2}\bigg[(1+\frac{\overline{\alpha}_2}{r_h^2})^3-1-\Lambda\overline{\alpha}_2+\frac{3\overline{\alpha}_2}{\beta (d-1)(d-2)}\\&\times\log{\bigg(1+\frac{4\beta Q^2}{r_h^{2d-4}}\bigg)}+\frac{24Q^2d\overline{\alpha}_2}{(d^2-4d+3)r_h^{2d-4}}F_1\bigg[1,\frac{d-3}{2(d-2)},\frac{3d-7}{2(d-2)},\frac{-4\beta Q^2}{r_h^{2(d-2)}}\bigg]\\&-\frac{12Q^2\overline{\alpha}_2}{\sqrt{\beta}(d-3)(d-4)r_h^{d-2}}\arctan{\bigg(\frac{2Q\sqrt{\beta}}{r_h^{d-2}}\bigg)}\bigg].\label{54}\end{split}
 \end{equation}  
 \begin{figure}[h]
 	\centering
 	\includegraphics[width=0.8\textwidth]{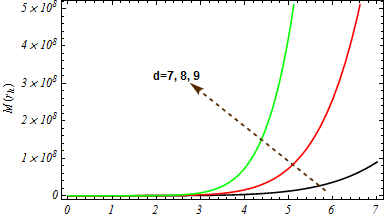}
 	\caption{The plots show how the mass $M$ (Eq. (\ref{54})) varies with horizon radius $r_h$. The chosen parameter values are $Q=0.5$, $\beta=0.3$, $\Sigma_{d-2}=100$, $\Lambda=-3$, and $\overline{\alpha}_2=0.5(d-3)(d-4)$.}\label{MLG1}
 \end{figure} 
 \begin{figure}[h]
 	\centering
 	\includegraphics[width=0.8\textwidth]{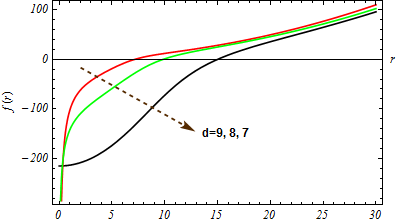}
 	\caption{Plots showing the behavior of $f(r)$ (Eq. (\ref{51})) in different dimensions. Particular values of the parameters are selected as $M=3\times10^8$, $Q=1.5$, $\beta=0.5$, $k=1$, $\Sigma_{d-2}=100$, $\Lambda=-3$, and $\overline{\alpha}_2=0.5(d-3)(d-4)$.}\label{Askar3d}
 \end{figure}
 \begin{figure}[h]
 	\centering
 	\includegraphics[width=0.8\textwidth]{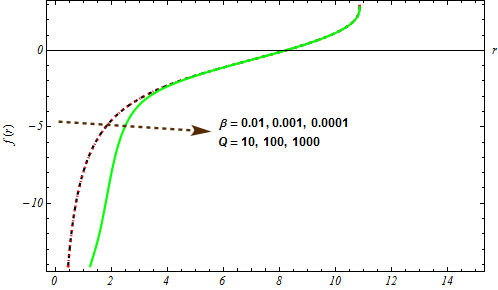}
 	\caption{Plots depicting the dependence of $f(r)$ (Eq. (\ref{51})) on $\beta$, and $Q$. Other parameters are defined as $M=3\times10^8$, $d=9$, $\Sigma_{d-2}=100$, $k=1$, $\Lambda=-0.03$ and $\overline{\alpha}_2=0.5(d-3)(d-4)$.}\label{Askar3Q}
 \end{figure}  
 The behavior of mass function dependent on horizon radius for suitably selected values of parameters $\beta$, $\overline{\alpha}_2$, $\Lambda$ and $Q$ is depicted in Fig. \ref{MLG1}. All those values of $r_h$ for which mass remains positive correspond to the horizons of the black hole. Similarly, the metric function (\ref{51}) for the mentioned values of these parameters in different spacetime dimensions is plotted in Figs. \ref{Askar3d} and Fig. \ref{Askar3Q}. The point at which the metric function (\ref{51}) vanishes for the fixed values of parameters $\beta$, $\Lambda$, $Q$, and $d$ refers to the horizon radius. 
 
 Using the above metric function, one can find Hawking temperature as
 \begin{eqnarray}\begin{split}
 T_H(r_h)&=\frac{r_h^6}{4\pi(r_h^4+2\overline{\alpha}_2r_h^2+\overline{\alpha}_2^2)}\bigg[-\frac{2\Lambda}{(d-2)r_h}+\frac{d-3}{r_h^2}+\frac{\overline{\alpha}_2(d-5)}{r_h^5}\\&+\frac{\overline{\alpha}_2^2(d-7)}{3r_h^7}+\frac{1}{\beta (d-2)r_h}\log{\bigg(1+\frac{4\beta Q^2}{r_h^{2d-4}}\bigg)}+\frac{4Q^2}{\sqrt{\beta}(d-2)(d-3)r_h^{d-1}}\\&\times\arctan{\bigg(\frac{2Q\sqrt{\beta}}{r_h^{d-2}}\bigg)}+\frac{8Q^2\big(Qd-Q-d^2+2d-3\big)}{(d^2-4d+3)(r_h^{2d-3}+4\beta Q^2r_h)}\bigg].\label{55}\end{split}
 \end{eqnarray} 
  \begin{figure}[h]
 	\centering
 	\includegraphics[width=0.8\textwidth]{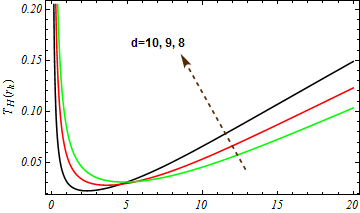}
 	\caption{The plots describe the changes in temperature $T_H$ (Eq. (\ref{55})) in different spacetime dimensions. The fixed values of the other parameters are considered as $Q=1.5$, $\beta=0.5$, $\Sigma_{d-2}=100$, $\Lambda=-3$, and $\overline{\alpha}_2=2(d-3)(d-4)$.}\label{temp3}
 \end{figure} 
 \begin{figure}[h]
 	\centering
 	\includegraphics[width=0.8\textwidth]{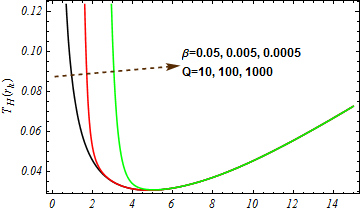}
 	\caption{The plots describe the changes in temperature $T_H$ (Eq. (\ref{55})) with different values of $Q$ and $\beta$. Furthermore, $d=10$, $\Sigma_{d-2}=100$, $\Lambda=-0.3$, and $\overline{\alpha}_2=2(d-3)(d-4)$ are considered as well.}\label{temp3q}
 \end{figure} 

The Hawking temperature of black hole as a function of $r_h$ is plotted in Figs. \ref{temp3} and \ref{temp3q}. It can be clearly understood from these plots that the smaller black holes are largely affected by the nonlinearity parameter $\beta$ and charge $Q$, however, for larger black holes the effects of these parameters on the Hawking temperature are very small. It is also shown that the temperature first decreases to its minimum value as $r_h$ approaches its critical value say $r_c$ but as $r_h$ crosses this value, temperature is increasing monotonically. Differentiation of Eq. (\ref{55}) yields
\begin{eqnarray}\begin{split}
\frac{dT_H}{dr_h}&=\frac{(2r_h^9+8\overline{\alpha}_2r_h^7+6\overline{\alpha}_2^2r_h^5)}{4\pi(r_h^4+2\overline{\alpha}_2r_h^2+\overline{\alpha}_2^2)^2}\mathcal{W}(r_h)+\frac{r_h^6}{4\pi (r_h^4+2\overline{\alpha}_2r_h^2+\overline{\alpha}_2^2)}\bigg(\frac{d\mathcal{W}}{dr_h}\bigg),\label{56}\end{split}
\end{eqnarray}  
where 
\begin{eqnarray}\begin{split}
\mathcal{W}(r_h)&=-\frac{2\Lambda}{(d-2)r_h}+\frac{d-3}{r_h^2}+\frac{\overline{\alpha}_2(d-5)}{r_h^5}+\frac{\overline{\alpha}_2^2(d-7)}{3r_h^7}+\frac{1}{\beta (d-2)r_h}\log{\bigg(1+\frac{4\beta Q^2}{r_h^{2d-4}}\bigg)}\\&+\frac{4Q^2}{\beta^{1/2}(d^2-5d+6)r_h^{d-1}}\arctan{\bigg(\frac{2Q\sqrt{\beta}}{r_h^{d-2}}\bigg)}+\frac{8Q^2\big(Qd-Q-d^2+2d-3\big)}{(d^2-4d+3)(r_h^{2d-3}+4\beta Q^2r_h)},\label{57}\end{split}
\end{eqnarray}
and
\begin{eqnarray}\begin{split}
\frac{d\mathcal{W}}{dr_h}&=\frac{2\Lambda}{(d-2)r_h^2}-\frac{3(d-3)}{r_h^2}-\frac{5\overline{\alpha}_2(d-5)}{r_h^6}-\frac{7\overline{\alpha}_2^2(d-7)}{3r_h^8}-\frac{8Q^2}{r_h^2(r_h^{2d-4}+4\beta Q^2)}\\&-\frac{1}{\beta (d-2)r_h^2}\log{\bigg(1+\frac{4\beta Q^2}{r_h^{2d-4}}\bigg)}-\frac{4Q^2(d-1)}{\beta^{1/2}(d^2-5d+6)r_h^{d}}\arctan{\bigg(\frac{2Q\sqrt{\beta}}{r_h^{d-2}}\bigg)}\\&+\frac{8Q^2\big(Qd-Q-d^2+2d-3\big)}{(d^2-4d+3)(r_h^{2d-3}+4\beta Q^2r_h)^2\big((2d-3)r_h^{2d-4}+4\beta Q^2\big)^{-1}}\\&-\frac{(2d-4)Q\sqrt{\beta}r_h^{d-3}}{r_h^{2d-4}+4\beta Q^2}.\label{58}\end{split}
\end{eqnarray}
One may find the entropy through Wald's method as 
\begin{equation}
S=2\pi(d-2)\Sigma_{d-2}\bigg(\frac{r_h^{d-2}}{d-2}+\frac{\overline{\alpha}_2r_h^{d-4}}{d-4}+\frac{\overline{\alpha}_2^2r_h^{d-6}}{3d-18}\bigg).\label{59}
\end{equation}
Finally, using the above expressions (\ref{56})-(\ref{59}) one can find the heat capacity as
\begin{eqnarray}\begin{split}
C_Q&=\frac{2\pi(d-2)\Sigma_{d-2}\mathcal{W}(r_h)\big(r_h^{d-3}+\overline{\alpha}_2r_h^{d-5}+\frac{\overline{\alpha}_2^2}{3}r_h^{d-7}\big)\big(r_h^{10}+2\overline{\alpha}_2r_h^8+\overline{\alpha}_2^2r_h^6\big)}{\big(2r_h^9+8\overline{\alpha}_2r_h^7+6\overline{\alpha}_2^2r_h^5\big)\mathcal{W}(r_h)+\big(r_h^{10}+2\overline{\alpha}_2r_h^8+\overline{\alpha}_2^2r_h^6\big)\big(d\mathcal{W}/dr_h\big)}.\label{60}\end{split}
\end{eqnarray}
\begin{figure}[h]
	\centering
	\includegraphics[width=0.8\textwidth]{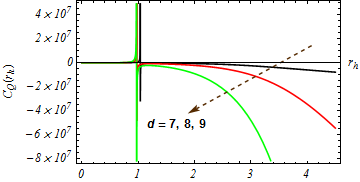}
	\caption{The plots describe the changes of specific heat $C_Q$ (Eq. (\ref{60})) in different spacetime dimensions. Additionally, $Q=1.5$, $\beta=0.5$, $\Sigma_{d-2}=100$, $\Lambda=-3$, and $\overline{\alpha}_2=2(d-3)(d-4)$ are selected as well.}\label{khan4}
\end{figure}
\begin{figure}[h]
	\centering
	\includegraphics[width=0.8\textwidth]{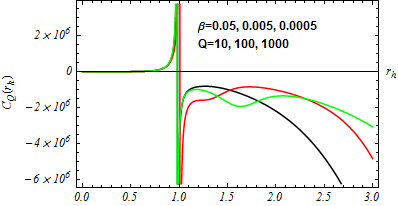}
	\caption{The plots describe the effects of $\beta$ and $Q$ on specific heat $C_Q$ (Eq. (\ref{60})). The other parameters are chosen as $d=8$, $\Sigma_{d-2}=100$, $\Lambda=-3$, and $\overline{\alpha}_2=2(d-3)(d-4)$.}\label{khan4q}
\end{figure}
\begin{figure}[h]
	\centering
	\includegraphics[width=0.8\textwidth]{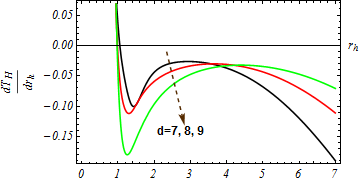}
	\caption{The plots describe the variations of $\frac{dT_H}{dr_h}$ (Eq. (\ref{56})) in different dimensions. Furthermore, $Q=10$, $\beta=0.05$, $\Sigma_{d-2}=100$, $\Lambda=-3$ and $\overline{\alpha}_2=2(d-3)(d-4)$.}\label{change3}
\end{figure}
\begin{figure}[h]
	\centering
	\includegraphics[width=0.8\textwidth]{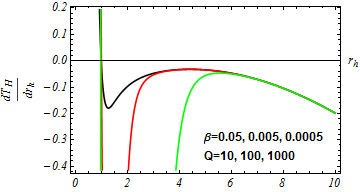}
	\caption{The plots describe the effects of $\beta$ and $Q$ on $\frac{dT_H}{dr_h}$ (Eq. (\ref{56})). The other parameters are considered as $d=9$, $\Sigma_{d-2}=100$, $\Lambda=-3$, and $\overline{\alpha}_2=2(d-3)(d-4)$ are selected as well.}\label{change3q}
\end{figure}

The black hole would be locally stable when $C_Q$ is positive. However, the zeros of this quantity correspond to the possibility of first-order phase transitions. One can identify the regions of positive heat capacity from Fig. \ref{khan4}. The effects of parameters $\beta$ and $Q$ on the stability of black holes can be seen in Fig. \ref{khan4q}. Additionally, one can see the roots of $dT_H/dr_h$ in Figs. \ref{change3} and \ref{change3q}. These roots are the singular points of heat capacity and present second-order phase transitions of a black hole.
\section{Summary and conclusion} 

 Analogous to the Einstein's theory of gravity, Lovelock gravity contains the metric derivatives of order two in its field equations. In addition to this, it can be viewed as the classical theory of gravity's high energy limit. As a result, Lovelock theories of any order attracted considerable interest, especially when it came to the search for new higher dimensional black holes. Here, I focused on the new Lovelock black hole solutions supported by double-logarithmic electrodynamics. The corresponding field equations are solved using this NLED model, and the Lovelock polynomial fulfilling these equations is then derived. Since the non-vanishing electric field does not provide the Lovelock polynomial in closed form, I took into consideration the scenario of a pure magnetic field in the chosen NLED model. Additionally, the Lovelock black holes' thermodynamic quantities associated with this polynomial equation (\ref{13}) are also calculated. I then proceeded to discuss the magnetized black holes of Einstein's theory, Gauss-Bonnet gravity and the third order Lovelock gravity in the background of double-logarithmic electrodynamics. The exact expressions for the metric function and important thermodynamic quantities for the black holes of these theories are derived as well. The behavior of the metric functions is also examined along with the thermodynamic quantities that were shown in each case. The thermodynamic quantities of the smaller black holes are demonstrated to be significantly affected by the double-logarithmic electromagnetic field, but as the size of these black holes increases, these effects get smaller and smaller.     
  
 It is important to note that when $\beta\rightarrow 0$, the resulting metric functions (\ref{27}), (\ref{39}) and (\ref{51}) describe the higher dimensional black holes with cosmological constant $\Lambda$ in Einstein-Maxwell, Gauss-Bonnet-Maxwell and third order Lovelock-Maxwell gravities, respectively. The derived metric functions, however, describe neutral black holes when $Q$ is set to zero. 
 
The impact of double-logarithmic electrodynamics on the phenomena of thermal fluctuations, critical behavior, quasi-normal modes, Hawking radiations, and greybody factors related to the Einsteinian and Lovelock black holes would be highly intriguing to study. Additionally, this NLED model can be utilized to analyze the Universe's acceleration.

\textbf{Data availability statement}
This manuscript has no associated data or the data will not be deposited.

\end{document}